\documentclass[twocolumn]{aastex631}

\usepackage{multirow}
\usepackage{amsmath}

\begin{document}

\title{A NIRCam-dark galaxy detected with the MIRI/F1000W filter in the MIDIS/JADES Hubble Ultra Deep Field}

\author[0000-0003-4528-5639]{Pablo G. P\'erez-Gonz\'alez}
\affiliation{Centro de Astrobiolog\'{\i}a (CAB), CSIC-INTA, Ctra. de Ajalvir km 4, Torrej\'on de Ardoz, E-28850, Madrid, Spain}

\author[0000-0002-5104-8245]{Pierluigi Rinaldi}
\affiliation{Kapteyn Astronomical Institute, University of Groningen, P.O. Box 800, 9700 AV Groningen, The Netherlands}

\author[0000-0001-8183-1460]{Karina I. Caputi}
\affiliation{Kapteyn Astronomical Institute, University of Groningen, P.O. Box 800, 9700 AV Groningen, The Netherlands}


\author[0000-0002-7093-1877]{Javier Álvarez-Márquez}
\affiliation{Centro de Astrobiolog\'{\i}a (CAB), CSIC-INTA, Ctra. de Ajalvir km 4, Torrej\'on de Ardoz, E-28850, Madrid, Spain}

\author[0000-0002-8053-8040]{Marianna Annunziatella}
\affiliation{Centro de Astrobiolog\'{\i}a (CAB), CSIC-INTA, Ctra. de Ajalvir km 4, Torrej\'on de Ardoz, E-28850, Madrid, Spain}

\author[0000-0001-5710-8395]{Danial Langeroodi}
\affiliation{DARK, Niels Bohr Institute, University of Copenhagen, Jagtvej 155, 2200 Copenhagen, Denmark}

\author[0000-0002-3305-9901]{Thibaud Moutard} 
\affiliation{European Space Agency (ESA), European Space Astronomy Centre (ESAC), Camino Bajo del
Castillo s/n, 28692 Villanueva de la Cañada, Madrid, Spain}

\author[0000-0002-3952-8588]{Leindert Boogaard}
\affiliation{Max Planck Institut f\"ur Astronomie, K\"onigstuhl 17, D-69117, Heidelberg, Germany}

\author[0000-0001-8386-3546]{Edoardo Iani}
\affiliation{Kapteyn Astronomical Institute, University of Groningen, P.O. Box 800, 9700 AV Groningen, The Netherlands}

\author[0000-0003-0470-8754]{Jens Melinder}
\affiliation{Department of Astronomy, Stockholm University, Oscar Klein Centre, AlbaNova University Centre, 106 91 Stockholm, Sweden}

\author[0000-0001-6820-0015]{Luca Costantin}
\affiliation{Centro de Astrobiolog\'{\i}a (CAB), CSIC-INTA, Ctra. de Ajalvir km 4, Torrej\'on de Ardoz, E-28850, Madrid, Spain}

\author[0000-0002-3005-1349]{G{\"o}ran {\"O}stlin}
\affiliation{Department of Astronomy, Stockholm University, Oscar Klein Centre, AlbaNova University Centre, 106 91 Stockholm, Sweden}

\author[0000-0002-9090-4227]{Luis Colina}
\affiliation{Centro de Astrobiolog\'{\i}a (CAB), CSIC-INTA, Ctra. de Ajalvir km 4, Torrej\'on de Ardoz, E-28850, Madrid, Spain}

\author[0000-0002-2554-1837]{Thomas R. Greve}
\affiliation{Cosmic Dawn Center (DAWN), Denmark}
\affiliation{DTU Space, Technical University of Denmark, Elektrovej, Building 328, 2800, Kgs. Lyngby, Denmark}
\affiliation{Dept.~of Physics and Astronomy, University College London, Gower Street, London WC1E 6BT, United Kingdom}

\author[0000-0001-7416-7936]{Gillian Wright}
\affiliation{UK Astronomy Technology Centre, Royal Observatory Edinburgh, Blackford Hill, Edinburgh EH9 3HJ, UK}

\author[0000-0001-6794-2519]{Almudena Alonso-Herrero}
\affiliation{Centro de Astrobiolog\'{\i}a (CAB), CSIC-INTA, Camino Bajo del Castillo s/n, E-28692 Villanueva de la Ca\~nada, Madrid, Spain}

\author[0000-0001-8068-0891]{Arjan Bik}
\affiliation{Department of Astronomy, Stockholm University, Oscar Klein Centre, AlbaNova University Centre, 106 91 Stockholm, Sweden}

\author[0000-0001-8582-7012]{Sarah E.~I.~Bosman}
\affiliation{Institute for Theoretical Physics, Heidelberg University, Philosophenweg 12, D–69120, Heidelberg, Germany}
\affiliation{Max-Planck-Institut f\"ur Astronomie, K\"onigstuhl 17, 69117 Heidelberg, Germany}

\author[0000-0003-2119-277X]{Alejandro Crespo G\'omez}
\affiliation{Centro de Astrobiolog\'{\i}a (CAB), CSIC-INTA, Ctra. de Ajalvir km 4, Torrej\'on de Ardoz, E-28850, Madrid, Spain}

\author[0000-0003-0589-5969]{Daniel Dicken}
\affiliation{UK Astronomy Technology Centre, Royal Observatory Edinburgh, Blackford Hill, Edinburgh EH9 3HJ, UK}

\author[0000-0001-6049-3132]{Andreas Eckart}
\affiliation{I.Physikalisches Institut der Universit\"at zu K\"oln, Z\"ulpicher Str. 77, 50937 K\"oln, Germany}

\author[0000-0003-4801-0489]{Macarena García-Marín}
\affiliation{European Space Agency, Space Telescope Science Institute, Baltimore, Maryland, USA}

\author[0000-0001-9885-4589]{Steven Gillman}
\affiliation{Cosmic Dawn Center (DAWN), Denmark}
\affiliation{DTU Space, Technical University of Denmark, Elektrovej, Building 328, 2800, Kgs. Lyngby, Denmark}

\author[0000-0001-9818-0588]{Manuel G\"udel}
\affiliation{Dept. of Astrophysics, University of Vienna, Türkenschanzstr 17, A-1180 Vienna, Austria}
\affiliation{ETH Zürich, Institute for Particle Physics and Astrophysics, Wolfgang-Pauli-Str. 27, 8093 Zürich, Switzerland}

\author[0000-0002-1493-300X]{Thomas Henning}
\affiliation{Max Planck Institut f\"ur Astronomie, K\"onigstuhl 17, D-69117, Heidelberg, Germany}

\author[0000-0002-4571-2306]{Jens Hjorth}
\affiliation{DARK, Niels Bohr Institute, University of Copenhagen, Jagtvej 155, 2200 Copenhagen, Denmark}

\author[0000-0002-2624-1641]{Iris Jermann}
\affiliation{Cosmic Dawn Center (DAWN), Denmark}
\affiliation{DTU Space, Technical University of Denmark, Elektrovej, Building 328, 2800, Kgs. Lyngby, Denmark}

\author[0000-0002-0690-8824]{\'Alvaro Labiano}
\affiliation{Telespazio UK for the European Space Agency, ESAC, Camino Bajo del Castillo s/n, 28692 Villanueva de la Ca\~{n}ada, Spain}

\author[0000-0001-5492-4522]{Romain A. Meyer}
\affiliation{Department of Astronomy, University of Geneva, Chemin Pegasi 51, 1290 Versoix, Switzerland}

\author[0000-0002-9850-2708]{Florian Pei$\beta$ker}
\affil{I.Physikalisches Institut der Universit\"at zu K\"oln, Z\"ulpicher Str. 77, 50937 K\"oln, Germany}

\author[0000-0002-0932-4330]{John P. Pye}
\affiliation{School of Physics \& Astronomy, Space Research Centre, Space Park Leicester, University of Leicester, 92 Corporation Road, Leicester LE4 5SP, UK}

\author[0000-0002-2110-1068]{Thomas P. Ray}
\affiliation{Dublin Institute for Advanced Studies, 31 Fitzwilliam Place, D02 XF86, Dublin, Ireland}

\author[0009-0003-6128-2347]{Tuomo Tikkanen}
\affiliation{School of Physics \& Astronomy, Space Research Centre, Space Park Leicester, University of Leicester, 92 Corporation Road, Leicester LE4 5SP, UK}

\author[0000-0003-4793-7880]{Fabian Walter}
\affiliation{Max Planck Institut f\"ur Astronomie, K\"onigstuhl 17, D-69117, Heidelberg, Germany}

\author[0000-0001-5434-5942]{Paul P.~van der Werf}
\affiliation{Leiden Observatory, Leiden University, P.O.~Box 9513, 2300 RA Leiden, The Netherlands}


\begin{abstract}

We report the discovery of {\it Cerberus}, an extremely red object detected with the MIRI Deep Imaging Survey (MIDIS) observations in the F1000W filter of the Hubble Ultra Deep Field. The object is detected at $S/N\sim6$, with $\mathrm{F1000W}\sim27$~mag, and undetected in the NIRCam data gathered by the JWST Advanced Deep Extragalactic Survey, JADES, fainter than the 30.0-30.5~mag $5\sigma$ detection limits in individual bands, as well as in the MIDIS F560W ultra-deep data ($\sim$29~mag, $5\sigma$). Analyzing the spectral energy distribution built with low-$S/N$ ($<5$) measurements in individual optical-to-mid-infrared filters and higher $S/N$ ($\gtrsim5$) in stacked NIRCam data, we discuss the possible nature of this red NIRCam-dark source using a battery of codes. We discard the possibility of {\it Cerberus} being a Solar System body based on the $<0\farcs016$ proper motion in the 1-year apart JADES and MIDIS observations. A sub-stellar Galactic nature is deemed unlikely, given that the {\it Cerberus'} relatively flat NIRCam-to-NIRCam and very red NIRCam-to-MIRI flux ratios are not consistent with any brown dwarf model. The extragalactic nature of {\it Cerberus} offers 3 possibilities: (1) A $z\sim0.4$ galaxy with strong emission from polycyclic aromatic hydrocarbons; the very low inferred stellar mass, $\mathrm{M}_\star=10^{5-6}$~M$_\odot$, makes this possibility highly improbable. (2) A dusty galaxy at $z\sim4$ with an inferred stellar mass $\mathrm{M}_\star\sim10^{8}$~M$_\odot$. (3) A galaxy with observational properties similar to those of the reddest little red dots discovered around $z\sim7$, but {\it Cerberus} lying at $z\sim15$, with the rest-frame optical dominated by emission from a dusty torus or a dusty starburst.
\end{abstract}


\section{Introduction} \label{sec:intro}

Building increasingly powerful telescopes operating at redder and redder wavelengths and (unexpectedly) discovering new types of galaxies at higher and higher redshifts is a classical industry now, established nearly 40 years ago. Indeed, deep optical surveys carried out in the late twentieth century using highly sensitive detectors were found to miss interesting galaxy populations of evolved and dust-rich galaxies at cosmological distances, both types being relatively bright at near-infrared wavelengths and easily detectable by new instruments, even though they were significantly less efficient in detecting photons and were affected by much larger backgrounds. The key for their discovery was large (red) colors (flux density ratios typically larger than an order of magnitude) between a near-infrared and an optical band, i.e., they were extremely red objects.

The first Extremely Red Objects (EROs) were reported using new ground-based near-infrared telescopes built nearly 40 years ago. They were found to be members of cosmologically relevant galaxy populations, comprising a variety of galaxies ranging from high redshift quasars to very young and dust-rich galaxies, or even $``$high$"$ redshift passive galaxies \citep[see, among many,][]{1988ApJ...331L..77E,2004ARA&A..42..477M,1996ApJ...471..720G,2003ApJ...587L..79F,2000MNRAS.317L..17P}.

With the development of very sensitive instruments onboard space telescopes such as {\it Spitzer}, the search was moved to longer wavelengths in the mid-infrared, at first compared with optical datasets \citep{2004ApJS..154..107W,2004ApJ...616...63Y}. 

Remarkably, when the {\it Hubble} Space Telescope gathered extremely deep data taken in the near-infrared up to 1.6~$\mu$m, with surveys such as CANDELS or the WFC3 Hubble Ultra Deep Field \citep{2011ApJS..197...36K,2011ApJS..197...35G,2013ApJS..209....3K}, {\it Spitzer}/IRAC data were still shown to detect unique distinctively red objects which were only visible at wavelengths longer than 3~$\mu$m, or at least were extremely faint in the near-infrared and optical spectral ranges. These so-called HST-dark galaxies were studied for around a decade, with significant uncertainties about their nature due to the strong sensitivity limitations of our multi-wavelength telescopes, even those operating in very different observational windows such as the (sub-)millimeter range \citep[see, e.g.,][]{2014ApJ...788..125S,2018A&A...620A.152F,2019ApJ...876..135A,2019Natur.572..211W}. JWST has started, from the earliest phases of the mission, to shed light on those HST-dark galaxies, i.e., mid-infrared bright near-infrared/optically faint sources which now constitute the bulk of the JWST galaxy exploration \citep{2023ApJ...946L..16P,2023ApJ...948L..18N,2023MNRAS.522..449B,2023arXiv231107483W,2023A&A...676A..26G}.

The ERO astrophysical industry is intimately tied to the discovery of submillimeter galaxies (SMGs, \citealt{1998Natur.394..241H}). Indeed, many EROs are dusty starbursts, which appear in mid- and far-infrared surveys in different flavors and with different names, starting from SMGs (see, e.g., \citealt{2002PhR...369..111B,2005ApJ...622..772C}) but also  (hot) dust-obscured galaxies (DOGs, \citealt{2008ApJ...677..943D,2010MNRAS.407.1701N}; hotDOGs, \citealt{2015ApJ...805...90T}), extremely red quasars (ERQ, \citealt{2017MNRAS.464.3431H}), or, more descriptive from an observational point of view, (ultra-)luminous infrared galaxies (U/LIRGs, \citealt{1996ARA&A..34..749S,2002A&A...384..848E,2005ApJ...630...82P,2005ApJ...632..169L,2007ApJ...660...97C,2009A&A...496...57M,2014ARA&A..52..415M}).


Last but not least, EROs, especially if selected at the faintest magnitudes and with appropriate bands for the color, overlap with very high-redshift galaxy samples selected using the Lyman break technique \citep{1996ApJ...462L..17S} at higher and higher redshifts \citep[e.g.,][]{2011ApJ...737...90B}. 

Remarkably, with the advent of JWST, both industries, searching for red objects and for those with higher and higher redshifts, have been merged at a profound level, with the possibility of finding double-break (Lyman and Balmer) red galaxies \citep{2023Natur.616..266L}, now known as little red dots \citep{2023MNRAS.519.3064F,2023A&A...677A.145U,2023arXiv230605448M,2023arXiv230514418B,2023arXiv230905714G,2023arXiv231107483W,2023arXiv231203065K,2024arXiv240108782P,2024arXiv240109981K}, or high redshift galaxies with very high equivalent width emission lines, first discovered with {\it Spitzer} \citep{2014ApJ...784...58S}, now appearing in many JWST works \citep{2023ApJ...946L..16P, Rinaldi_23a}. Interestingly, some of these sources have been confirmed to present broad emission line components possibly linked to an active galactic nucleus (AGN; \citealt{2023ApJ...954L...4K,2023arXiv230905714G}).

The interesting subtopics of finding and characterizing dusty galaxies on the one hand, and high redshift galaxy candidates on the other, has been demonstrated to be very relevant for JWST surveys, which can easily mis-identify one type of galaxy with the other \citep[see, e.g.,][]{2023ApJ...943L...9Z,2022arXiv220802794N}.

The MIRI instrument onboard the JWST \citep{2015PASP..127..584R,2023PASP..135d8003W} now offers the possibility to search for red galaxies with longer color baselines, beyond the wavelengths first probed by {\it Spitzer}/IRAC and now easily accessible with impressive depths ($\sim30$~mag and beyond) using JWST/NIRCam. Even though MIRI cannot reach NIRCam depths, very deep imaging at 5.6\,--\,12.8~$\mu$m reaching magnitudes around 27\,--\,29 is already available and opens a whole new window for the search and characterization of high redshift galaxies \citep{2023ApJ...949L..18P,2023arXiv230514418B,Rinaldi_23a,2024arXiv240108782P}. 

In this paper, we report the discovery of a  MIRI-bright red object appearing in the MIRI Deep Imaging Survey (MIDIS) F1000W imaging of the Hubble Ultra Deep Field (\"Ostlin et al., 2024, in prep.). The source is extremely faint at NIRCam long wavelengths, i.e., it is a NIRCam-dark object lying at the edge of the detection limit ($\sim30.5$~mag) of one of the deepest JWST surveys available to date, JADES \citep{2023arXiv230602465E}, having not been catalogued before \citep{2023ApJS..269...16R,2023arXiv231012340E}. Consequently, our object  presents an extremely red color between 4.4 and 10~$\mu$m, and, in fact, the source is also red with respect to the bluest MIRI filter, as revealed by the MIDIS F560W imaging. 

This letter is organized as follows. Section~\ref{sec:data} presents the deep MIRI data used in this letter, as well as ancillary observations taken with NIRCam, ALMA, and VLT. Section~\ref{sec:selection} describes our method to detect NIRCam-dark galaxies, and the case for a positive detection in the Hubble Ultra Deep Field, the source we call {\it Cerberus}. Section~\ref{sec:sed} presents the spectral energy distribution of {\it Cerberus} and discusses the several possible interpretations, analyzing the implied physical properties for each one of them. Finally, Section~\ref{sec:conclusions} presents a summary of our findings.

Throughout the paper, we assume a flat cosmology with $\mathrm{\Omega_M\, =\, 0.3,\, \Omega_{\Lambda}\, =\, 0.7}$, and a Hubble constant $\mathrm{H_0\, =\, 70\, km\,s^{-1} Mpc^{-1}}$. We use AB magnitudes \citep{1983ApJ...266..713O}. All stellar mass and SFR estimations assume a universal \citet{2003PASP..115..763C} initial mass function (IMF), unless stated otherwise.

\section{The MIRI deep F1000W survey in the HUDF, and ancillary data}
\label{sec:data}

\subsection{MIRI data}

MIRI data in the F1000W filter were taken as part of the MIRI Deep Imaging Survey (MIDIS, \"Ostlin et al. 2024, in prep.; PID 1283) in December 2023. The observation consisted of 11 exposures, each with 100 groups, FASTR1 readout, and 10 integrations, for a total on source exposure time of 30800~s, centered on the Hubble Ultra Deep Field (HUDF). The dithering pattern was set to large-size cycling, with the 11 exposures taken in different positions on the sky separated by up to 10\arcsec. 


The reduction of the MIRI data followed the methodology described in \citet{2024arXiv240108782P}, based on the official {\sc jwst} pipeline (v1.13.4, jwst\_1202.pmap) and offline bespoke procedures. Briefly, a super-background strategy is used to build background maps for each single image using all the other exposures (since they were taken during the same campaign), which results in a very homogeneous background in terms of level and noise. Quantitatively for the data used in this letter,  the standard deviation of the background pixels for each cal.fits file obtained with the standard official pipeline is reduced by a factor of $\sim3.5$ by using our super-background method, and the depth in the final mosaic improves by 0.8~mag. We refer the reader to Appendix~A in \citet{2024arXiv240108782P} for more details. Known sources are masked to avoid biasing the determination of the very local background in the super-background image. Our F1000W final mosaic, reduced with a 60 milliarcsec pixels, presents an average 5$\sigma$ depth of 26.8~mag, measured in a 0\farcs4 radius circular aperture, and taking into account noise correlation (linked to the drizzling method used in the mosaicking) and after applying the aperture correction (a 1.3 factor). 

\begin{figure*}[htp!]
\centering\includegraphics[width = \textwidth]{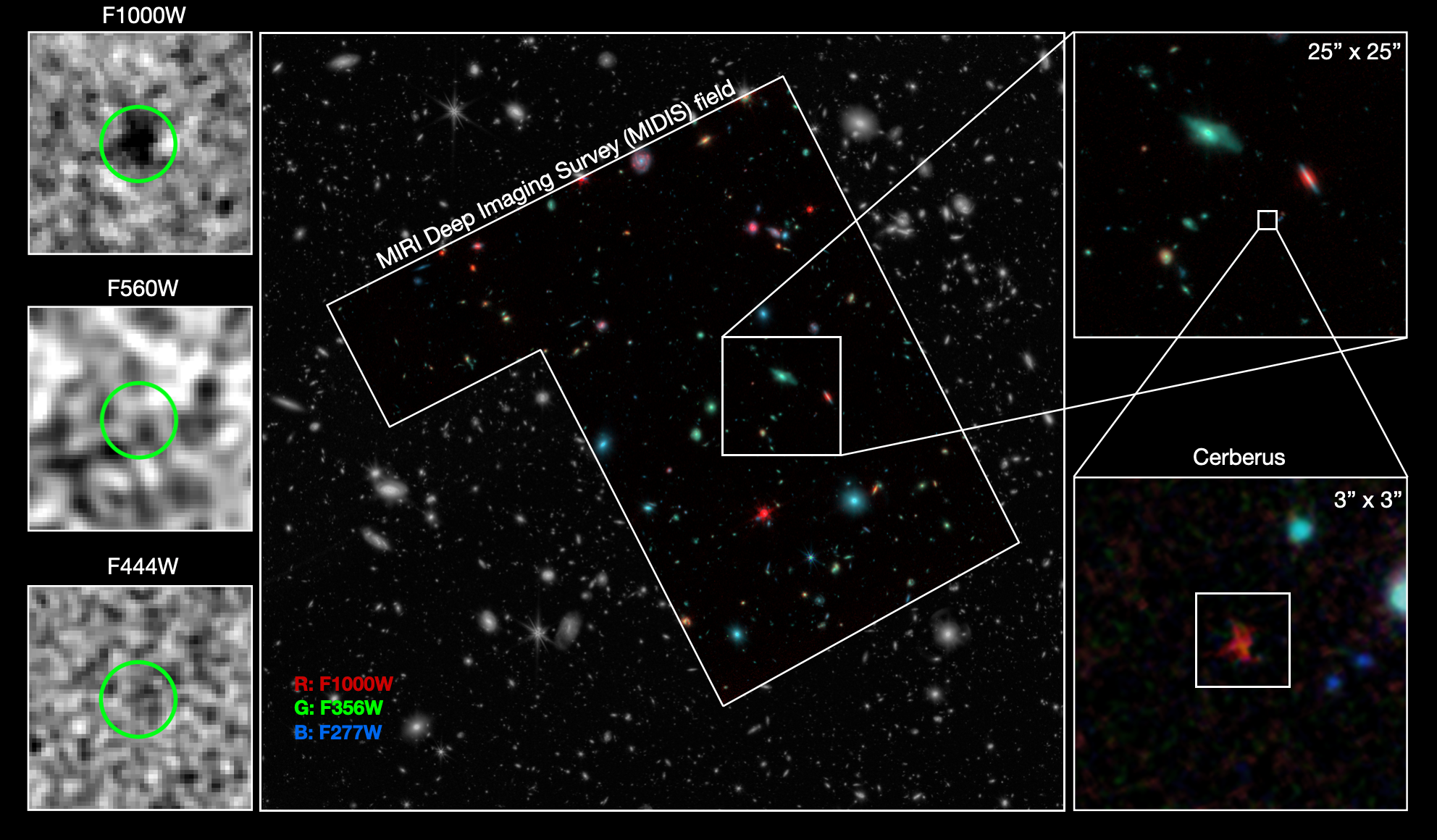}
\caption{In the center of the figure, we show an RGB composition of the MIDIS F1000W field. The color images have been built with JADES data in 2 NIRCam filters, F277W and F356W, and the MIDIS MIRI F1000W filter (all images convolved to the same PSF of MIRI/F1000W). In the background, we show the HUDF JADES data in grayscale. We provide a series of zoomed-in RGB frames that lead to the NIRCam-dark MIRI source, {\it Cerberus}, whose image is located at the bottom right corner. The NIRCam-dark MIRI source {\it Cerberus} is highlighted with a square in that RGB frame. Additionally, we provide postage stamp images (2\arcsec$\times$2\arcsec) of {\it Cerberus} in MIRI/F1000W, MIRI/F560W, and NIRCam/F444W (left column), showcasing that {\it Cerberus} (marked with an r=0\farcs25 green circle) is clearly detected only at 10~$\mu$m.}
\label{fig:midis}
\end{figure*}

We present the MIDIS F1000W data in Figure~\ref{fig:midis} with an RGB image built in combination with some NIRCam long wavelength data. 

\subsection{Ancillary data}

In this subsection, we describe other datasets in the HUDF that we have used to complement the MIRI data.

\subsubsection{Other MIRI data}

For this letter, we combined the recently acquired MIDIS F1000W data with the F560W ultra-deep observations also carried out by MIDIS in December 2022. These data will be described in \"Ostlin et al. (2024, in prep.) and consist of $\sim40$~hours on source taken in the HUDF, which allowed us to reach 28.7~mag 5$\sigma$ for point-like sources measured in an r=0\farcs23 circular aperture (the radius being chosen to ensure a $\sim70$\% encircled energy).

\subsubsection{NIRCam data}

The MIRI/F1000W data were also complemented with NIRCam imaging taken by the JWST Advanced Deep Extragalactic Survey, JADES \citep{2023arXiv230602465E}, Data Release 2  (\citealt{2023arXiv231012340E}), which includes observations from the JWST Extragalactic Medium-band Survey (JEMS, \citealt{2023ApJS..268...64W}) and the First Reionization Epoch Spectroscopically Complete Observations (FRESCO, \citealt{2023MNRAS.525.2864O}). This dataset provides a total of 14 bands from 0.9 to 4.8~$\mu$m (6 at short-wavelength, SW, and 8 at long-wavelength, LW), with 5$\sigma$ depths ranging from 30.5 to 30.9~mag (measured in a 0\farcs2 radius circular apertures). We remark that JADES is one of the deepest NIRCam surveys on the sky, only matched in depth (in some bands) by the MIDIS NIRCam-parallel project \citep{2023ApJ...951L...1P} and The Next Generation Deep Extragalactic Exploratory Public Near-Infrared Slitless Survey, NGDEEP \citep{2023arXiv230205466B}, which means that our search for NIRCam-dark galaxies is really probing the current depth limitations of JWST extragalactic surveys. 

\subsubsection{ALMA data}

The Atacama Large Millimeter/submillimeter Array (ALMA) Spectroscopic Survey in the Hubble Ultra Deep Field (ASPECS) is a Cycle 4 Large Program over a 4.6 arcmin$^{2}$ scan at 1.2~mm \citep{Decarli_2020, gonzalez-lopez2020} and 3.0~mm \citep{Decarli_2019, gonzalez-lopez2019}. The ultra-deep 1.2\,mm data reaches an rms sensitivity of 9.3\,$\mu$Jy/beam, with beam dimensions of $1\farcs5\times1\farcs1$.  The 3.0\,mm data reaches 1.4\,$\mu$Jy/beam, with a $1\farcs8\times1\farcs5$ beam.


\subsubsection{MUSE data}

The HUDF has been extensively studied in the last decade with the Multi Unit Spectroscopic Explorer \citep[MUSE,][]{2010SPIE.7735E..08B} mounted on the Very Large Telescope (VLT) as part of the MOSAIC and UDF-10 fields (GTO programs 094.A-0289(B), 095.A-0010(A), 096.A-0045(A) and 096.A-0045(B), PI: R.~Bacon) and the most recent MXDF observations (GTO Large Program 1101.A-0127, PI: R.~Bacon). The MUSE data has a spectral wavelength range between 4700 -- 9300\,{\AA} and a spectral resolving power (R) that ranges from 1770 (4800\,{\AA}) to 3590 (9300\,{\AA}). The HUDF dataset was taken during 140 hours on-source with the VLT Adaptive Optics Facility (AOF) and its GALACSI adaptive optics module, as detailed in \citet{2016SPIE.9909E..2SK} and \citet{2018SPIE10703E..02M}, achieving a spatial sampling of 0\farcs2 $\times$ 0\farcs2, with strong variations of up to a factor of $\times4$. More details about these programs can be found in \citet{Bacon_2017, 2023A&A...670A...4B}.






\section{Selection and validation of NIRCam-dark sources}
\label{sec:selection}

This letter presents a source that was discovered on the MIDIS F1000W image and found not to be included in existing NIRCam published catalogs \citep{2023ApJS..269...16R,2023arXiv231012340E}. In this section, we describe the selection procedure for such types of NIRCam-dark sources.

\begin{figure*}[htp!]
\centering
\includegraphics[width=0.9\textwidth]{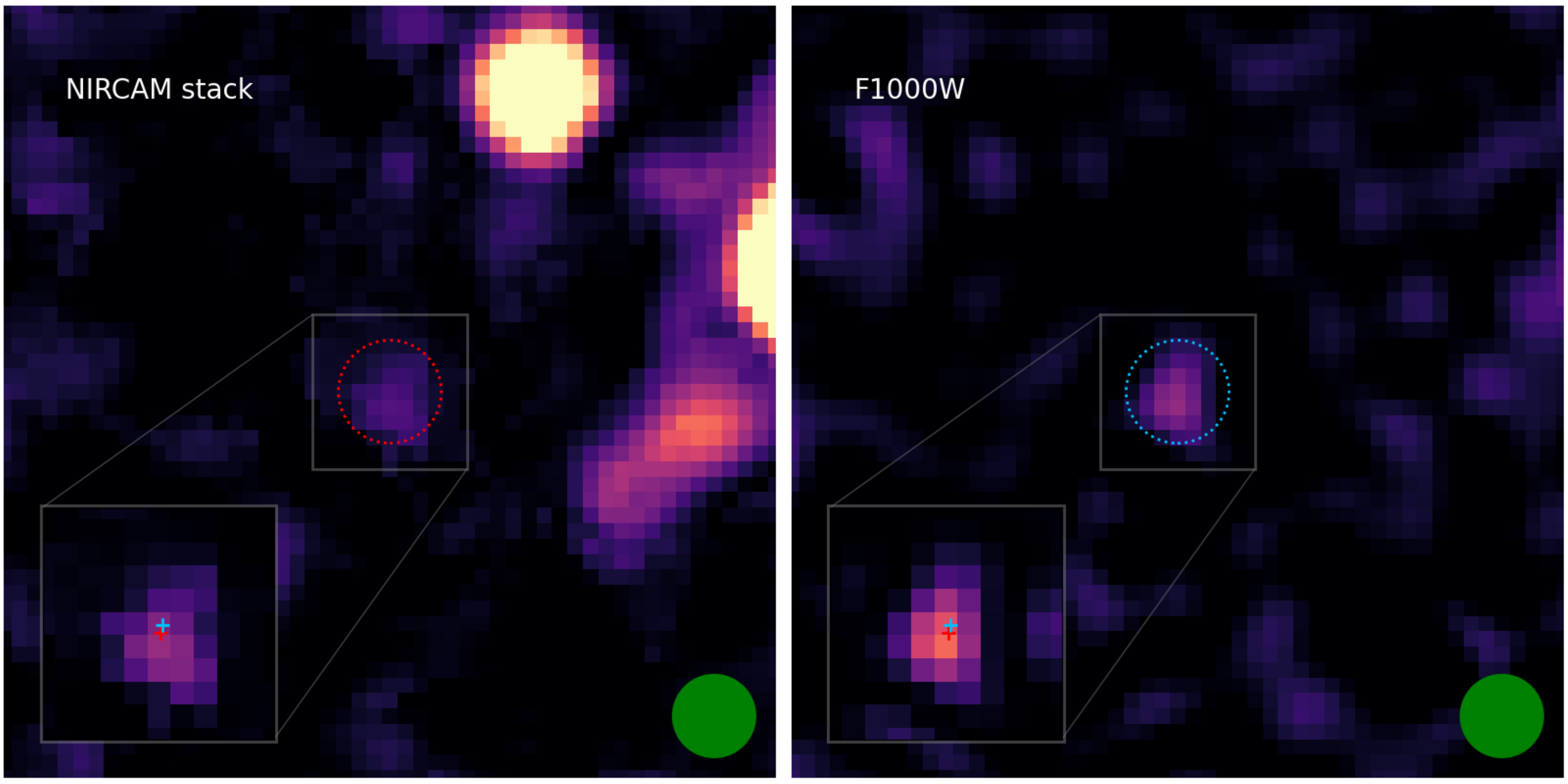}
\caption{Stacked NIRCam (left panel) and MIRI/F1000W (right panel) images of the region around {\it Cerberus}. The stacked NIRCam image was produced from individually PSF matched LW images. Note that the spurious pixels in the F460M image (see text for details) are removed in the median stacking procedure. The angular size of the images and zoomed inset is 3\arcsec$\times$3\arcsec and 0\farcs6 $\times$0\farcs6, respectively. The red (blue) circle shows the location of the detected source in F1000W and has a radius of 0\farcs2. The green filled circle shows the resolution of the F1000W image (with a diameter of 0\farcs328). In the zoomed insets, we also show the centroid positions with the blue and red crosses, corresponding to the F1000W image and the NIRCam stack, respectively.}
\label{fig:centroids}
\end{figure*}

\subsection{Selection of NIRCam-dark sources}

In order to select NIRCam-dark sources, we executed the \textsc{sextractor} software \citep{1996A&AS..117..393B} directly  using the MIRI/F1000W as the detection image. We followed the hot-mode extraction described in \citet{2013ApJS..206...10G}, which is particularly effective for detecting extremely faint sources. 

Once we had a hot MIRI/F1000W catalog, we anti cross-matched it (using a 0\farcs3 search radius) with the official JADES DR2 catalog published by \citet[][which superseded the DR1 catalog]{2023arXiv231012340E}, constructed by selecting sources in a LW stack. The goal of the anti cross-match was to isolate sources uniquely detected by MIRI/F1000W. The result was a list of 17 \textit{potential} NIRCam-dark sources selected at 10 $\mu$m.

We meticulously examined each of the \textit{potential} NIRCam-dark sources in the final mosaic to rule out the presence of an unflagged cosmic ray or any kind of hot and/or bad pixels that could have caused a spurious detection.


Visual inspection of the individual exposures revealed that for 16 of the 17 candidates, there was indeed a hot pixel in one of the 11 exposures which had survived the flagging and stacking process performed to produce the final mosaic. We automated and generalized the source validation methodology by producing stacked mosaics with a limited number of exposures, excluding 1, 2, or 3 of the full set randomly. This procedure, obviously, affects the depth. Explaining the latter case in more detail as an example, we started from our collection of 11 \texttt{*\_cal.fits} files, and we created the 165 unique mosaics which can be obtained by excluding sets of 3 \texttt{*\_cal.fits} files for each partial mosaic. After building the 165 realizations, we inspected the 17 \textit{potential} NIRCam-dark sources,  concluding that only one source consistently appeared in all 165 different partial mosaics. The remaining candidates were found to vanish in some of the realizations, in fact even in the mosaics built excluding one single exposure, indicating their most-likely spurious nature. To further validate these findings, we made an in-depth analysis of each \texttt{*\_cal.fits} file, confirming the presence of unflagged hot and/or bad pixels in some of them that indeed led to false detections in a number of the realizations. 

One source of the 17 pre-selected candidates did not vanish in any of the partial mosaics, even when removing 3 random images. For this source, we measured photometry in a small 0\farcs25 radius aperture (which maximized $S/N$ on the partial mosaics), as well as the sky noise in a square 3\arcsec$\times$3\arcsec\, region around, on all the mosaics produced by excluding 1, 2, and 3 exposures. We found consistent photometry and a degradation of the $S/N$ consistent with the differences in exposure time. For example, for the skip-3 mosaics, we found an average magnitude 27.05$\pm$0.19~mag, consistent with our measurement in the full mosaic (see Section~\ref{sec:sed}).

However, we found 2 contiguous exposures, numbers 7 and 8, which, if removed, produced a dimming of the source to 27.4~mag, still within 2$\sigma$ of the measurement using the full dataset. Each one of these exposures has an integration time of 2800~s, with the start of the first exposure and the end of the second separated by $\sim$5700~s. We examined these exposures in detail and found no signs of a cosmic ray hit, but a slightly enhanced signal in an extended region. This could be due to the general properties of the detector (which consistently showed a dimmer region in the super-background frames for all images). It could also be due to a cosmic ray shower, which are known to affect MIRI imaging observations \citep{2024arXiv240316686D}, with an incidence of several per exposure. Cosmic ray showers are extended persistence artifacts from cosmic ray hits and typically affect the detector for several minutes \citep{2023PASP..135d8003W}. However, a high flux cosmic ray hit could cause a shower lasting long enough to enhance the signal seen in these exposures. We also note that the enhancement in signal is stronger in exposure 7 compared to 8, which is also consistent with persistence \citep[see][]{2023A&A...672A.108A}. Given that the two relatively (and slightly) brighter images were contiguous, we do not rule out a transient behavior of the source. Based on the perseverance of the signal from the source in all partial stacks of the MIRI F1000W dataset, we concluded that it was not spurious.

The NIRCam-dark source we identified with this methodology was dubbed {\it Cerberus}, to acknowledge that it is a weird beast lurking in the darkest corners of the JWST surveys and with different interpretations/heads for its nature, as will be shown in Section~\ref{sec:sed}. Its coordinates (J2000) are $\alpha =$ 03:32:38.09 (hours), $\delta = $$-$27:47:11.81 (degrees). NIRCam and MIRI images of {\it Cerberus} are presented in Figure~\ref{fig:midis}.



\subsection{Multi-wavelength validation and observational properties of the NIRCam-dark object {\it Cerberus}}
\label{sec:validation}

In order to further validate and characterize {\it Cerberus}, we searched for detections by forcing photometric measurements in the datasets described in Section~\ref{sec:data}, more specifically in the ultra-deep observations (among or even the deepest on the sky) provided by JWST/NIRCam, ALMA, and VLT/MUSE. 

\subsubsection{JWST/NIRCam measurements}

Directly linked to our selection criteria, the NIRCam-dark {\it Cerberus} object was not present in any of the catalogs of the JADES NIRCam survey \citep{2023ApJS..269...16R,2023arXiv231012340E}. We then proceeded to force photometric measurements in all bands (described in Section~\ref{sec:sed}). Forced photometric measurements in all NIRCam individual bands do not provide any measurement signal-to-noise ratio $S/N>5$.  

We then stacked NIRCam images to search for faint emission. We constructed two different sets of stacks. One set consisted in cumulative stacks constructed by averaging NIRCam bands starting from the reddest and from the bluest. Another set consisted  of disjoint stacks (to avoid noise spikes in one band affecting the measurements in cumulative stacks), combining 2, 3, 4 and 5 bands, always starting from the reddest and not repeating them in different stacks of the same set. 

Figure~\ref{fig:centroids} shows a stacked NIRCam image (including all filters at $\lambda>3\mu$m) convolved to the resolution of the F1000W data (FWHM=0\farcs328). Each calibrated individual NIRCam filter was rebinned to a 0\farcs06 pixel scale and PSF-matched to F1000W using WebbPSF \citep[v.~1.2.1][]{Perrin2014} models and the method described in \citet{Melinder-2023}, and then median-stacked. We measured the centroid of the source for both JWST instruments (using \texttt{photutils} with a circular footprint of 3 pixels radius), finding an agreement within 0.26 pixels, i.e., 0\farcs016. Remarkably, the source is slightly extended to the NW in both images. The morphology of the source is similar for the NIRCam stack and the MIRI F1000W image, with a tadpole-like shape extending to the upper right, confirming a detection of the source with NIRCam stacked data (see next section).
We also notice a qualitative resemblance between the structure of {\it Cerberus} and the $z=10.6$ galaxy GN-z11 \citep{Oesch-2016, Tacchella-2023}, with a concentrated flux distribution identified with the core of the galaxy and an extended haze component.

\subsubsection{MUSE measurements}


We extracted the MUSE 1D spectrum at the coordinates of our target within circular apertures of different sizes (0\farcs4 and 0\farcs8 radius) but found no significant spectral feature at the position of the NIRCam-dark candidate. This is also confirmed by the publicly available catalog of detected sources by \citet{2023A&A...670A...4B}.



\subsubsection{ALMA measurements}

The source is not detected at 1.2\,mm and 3.0\,mm. This finding implies $5\sigma$ upper limits on the flux density at 1.2\,mm and 3.0\,mm of 47\,$\mu$Jy and 7\,$\mu$Jy respectively, assuming it is effectively an unresolved point-like source.

  


\section{Spectral energy distribution analysis of {\it Cerberus}}
\label{sec:sed}

\subsection{Multi-wavelength Photometry}

A spectral energy distribution (SED) for {\it Cerberus} was built by measuring photometry in all available NIRCam (including both individual bands and stacks) and MIRI data,  forcing measurements at the position determined for {\it Cerberus} with the MIRI data. Given the faint and small nature of our source, we measured forced photometry in several circular apertures with radii ranging from 0.1\arcsec\, to 0.6\arcsec. We measured the background in a small square region around {\it Cerberus} (3\arcsec\, on the side) and calculated the noise using non-adjacent pixels (5 pixels apart) to take into account noise correlation, as explained in \citet{2023ApJ...951L...1P}, obtaining similar depth results (within 0.2~mag) to those reported in \citet{2023arXiv230602465E} for the NIRCam bands. Photometric measurements are provided in Table~\ref{tab:photometry}.

None of the measurements in individual NIRCam bands have a $>5\sigma$ significance, which we use as our limit to define a detection in each filter.

Given our magnitude measurement in the F1000W band, $27.13\pm 0.17$~mag, obtained with photometry in an aperture of radius 0\farcs3, which maximizes the $S/N$, and the $S/N\sim3$ measurements in individual bands and $S/N>5$ in  stacked data for the reddest LW NIRCam channels, we infer a very red color, $\mathrm{F444W-F1000W}=3.8\pm$0.4~mag. Using the reddest disjoint stack with an actual detection (i.e., $S/N>5$), we obtain a color $[4.497]-[10.0]=2.8\pm0.3$~mag.

These measurements indicate that {\it Cerberus} is not robustly detected in any individual NIRCam filter, but consistent with being slightly below the 5$\sigma$  detections limits in some bands, quoted to be around 30--31~mag for the JADES DR2 data \citep{2023arXiv230602465E}. We remark that the F460M filter, which provides the brightest magnitude, presents a group of 4 pixels to the North of the position of our galaxy, which we confirmed to be an artifact of the reduction, and it is not seen in any other band (even the ones with deeper limiting magnitudes). We excluded this filter from the rest of analysis.

We measured photometry in the stacked data in the same apertures used for the individual bands, assigning average aperture corrections for the filters included in each stack. The 5 LW stacks with the highest number of bands all provide $S/N>5$ fluxes (the average being $S/N\sim6$), ranging from 29.9 to 30.8~mag. This means that the (low-S/N) SED is characterized by a red color extending through all the $S/N>5$ stacks from 2 to 4~$\mu$m and beyond, up to $29.9\pm$0.2~mag for the first (reddest) stack (including F480M$+$F444W), which reaches $\sim5$~$\mu$m. For the SW stacks, no significant signal is recovered in any of them. We measure a $5\sigma$ upper limit of $\sim31$~mag for the stack adding all SW data. The comparison between SW and LW stacks implies a $\gtrsim$0.6~mag color jump from the $\sim1-2$ to the $\sim3-5$~$\mu$m ranges.

The SED of {\it Cerberus} is, in fact, very similar to the little red dots (LRDs) recently presented and characterized in a number of papers (e.g., \citealt{2023Natur.616..266L,2023arXiv230607320L}, \citealt{2023arXiv230605448M}, \citealt{2023arXiv230514418B}, \citealt{2023arXiv231107483W},  \citealt{2024arXiv240108782P}, \citealt{2024arXiv240109981K}), but shifted to longer wavelengths. Based on detections of {\it Cerberus} in stacked images, we can infer that the $\mathrm{F277W-F444W}>1$~mag general selection criterion of LRDs, and, more specifically, the 3~mag color of the example source 203749 in \citet{2024arXiv240108782P}, compares well, if we multiply the observed wavelengths by a factor of $\sim$2, with the $\mathrm{F444W-F1000W}\sim3.5$~mag, or, using the F560W $5\sigma$ upper limit, $\mathrm{F560W-F1000W}\gtrsim2$~mag. The flat F150W$-$F200W color distinctive of LRDs (the largest value allowed by the selection is typically 0.5~mag), i.e., a change in slope at around 0.3-0.4~$\mu$m in the rest-frame for the typical redshift of LRDs $5<z<9$, compares well with the color between 3 and 4~$\mu$m for {\it Cerberus}, just multiplying again the observed wavelengths by a factor of $\sim$2. We remark that this comparison is based on S/N$>$5 measurements in stacked data. That rest-frame UV color is indeed consistent with the reddest LRDs in \citet{2024arXiv240108782P} or the   LRD of which  a NIRSpec spectrum was presented in \citet{2023arXiv231203065K}. The analogy can also be established in terms of the apparent magnitude: the typical magnitude of LRDs in the blue part of their SED lies around 27-28~mag, {\it Cerberus} being  $\gtrsim10$ times dimmer in NIRCam LW bands. We note that the difference in luminosity distance between $z\sim7$, the median redshift of LRDs \citep{2024arXiv240108782P} and $z\sim14$ (a factor of 2 larger, see also Section~\ref{sec:photoz}) will translate to a flux ratio between bands covering the same rest-frame wavelength around an order of magnitude for the same type of object.


\begin{figure}[htp!]
\centering
\includegraphics[clip,trim=36.8cm 5.0cm 0.0cm 4.0cm,width=8.cm,angle=0]{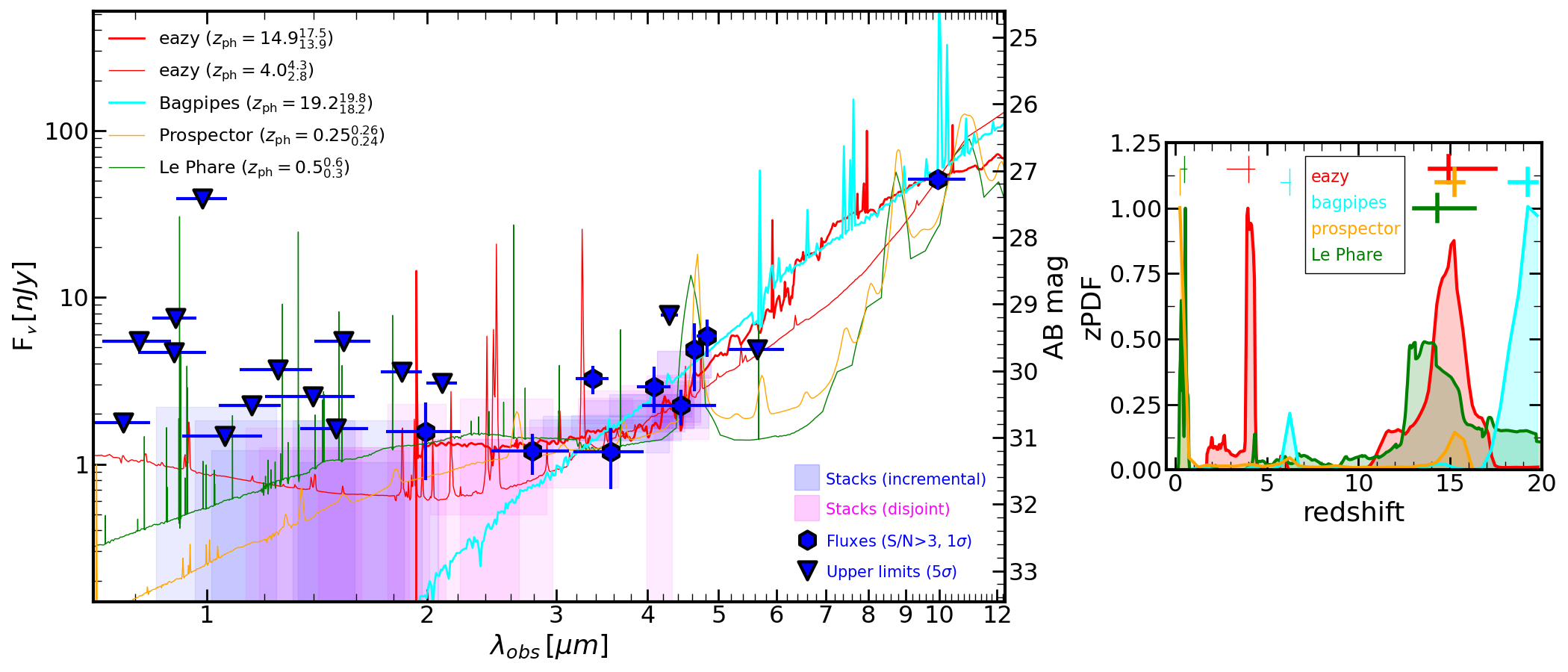}
\caption{Photometric redshift probability distribution functions of {\it Cerberus}. All distributions are normalized to unity at the peak. We show the main solutions and also secondary solutions at $z<10$ and $z>10$. Different codes were used in the SED analysis, described in Appendix~\ref{appA}.}
\label{fig:photoz}
\end{figure}

\subsection{Analysis of the SED of {\it Cerberus}}

In the following subsections, we discuss possible interpretations of the SED of {\it Cerberus}. The NIRCam and MIRI datasets were taken more than one year apart. However, given the spatial coincidence of the object in the NIRCam and MIRI observations (centroids separated by 0\farcs016), we proceed by assuming that they are the same object and not a chance alignment of two different objects.

\subsubsection{Extragalactic origin}
\label{sec:photoz}

The SED of {\it Cerberus} was fitted to estimate a photometric redshift and obtain its physical properties in a 2-step procedure. We used several codes for this purpose, separating the two tasks to analyze facets of each method and understand the (possibly systematic) uncertainties in detail. We used the fluxes described above as well as the ALMA upper limits quoted in Section~\ref{sec:data}, shown to be important for high redshift sources (even for determining their redshift, see \citealt{2024A&A...681L...3M}), and particularly for an ERO such as {\it Cerberus}. We provide details about the fitting methods in Appendix~\ref{appA}.

\begin{figure*}[htp!]
\centering
\includegraphics[clip,trim=0.0cm 0.0cm 0.0cm 0.0cm,width=18.cm,angle=0]{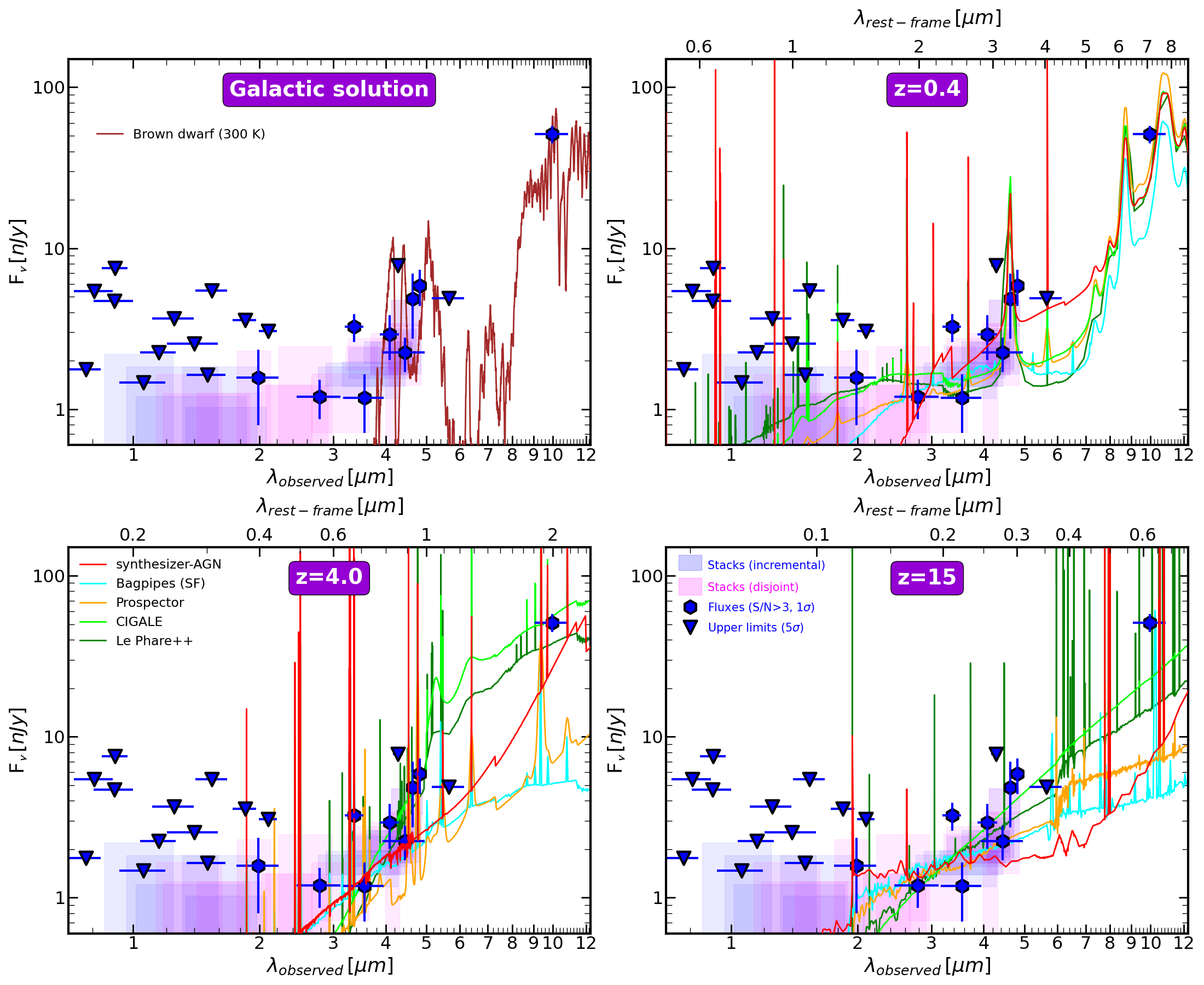}
\caption{Detailed analysis of the spectral energy distribution of {\it Cerberus} for the main redshift solutions. The symbols are the same as described in Figure~\ref{fig:photoz}. We provide best-fitting models provided by different codes that include both stellar and AGN emission. 
The top-left panel shows the best-fitting brown dwarf template from the ATMO2020++ models, for a Y-type dwarf with an effective temperature of $\sim 300$K at a distance of $\sim 300$ pc. 
The rest of the panels show results for $z=0.4$ (top-right), $z=3.5$ (bottom-left), and $z=15$ (bottom-right).}
\label{fig:sps}
\end{figure*}

The results of our SED modeling analysis are given in Figures~\ref{fig:photoz} (photometric redshift probability distributions) and ~\ref{fig:sps} (SED models).

We remark that the SED of {\it Cerberus} is very noisy, given that it is an ultra-faint source at the very depth edge of the JWST surveys. Therefore, the physical properties of the source are highly uncertain, starting from its redshift (see several peaks in Figure~\ref{fig:photoz}).

However, leaving the redshift apart, there are two physical properties for which our results are similar among all SED modelling codes, mainly because they are linked to the 2 main observational characteristics of {\it Cerberus}, i.e., its faintness (translating to relatively low luminosity) and red color. The very red color of our NIRCam-dark object can only be reproduced with significant amounts of dust, either linked to PAHs if the redshift is low, $z\sim0.4$, or warm/hot dust if the redshift is around $z\sim4$ or $z\sim15$. For the lowest redshift, all codes provide very low stellar masses, around $\sim10^{5-6}$~M$_\odot$. The very small stellar mass, relatively high dust attenuation, and strong PAH emission is an uncommon combination, at odds with the mass-metallicity relationship \citep[e.g.,][]{2004ApJ...613..898T} and the weakening of PAH features at low metallicities \citep[e.g.,][]{2005ApJ...628L..29E}. For the higher redshift solutions, $z\sim4$ and $z\sim15$, all codes provide masses around $\sim10^{7-8}$~M$_\odot$, and they all need large warm dust contents, which can be explained with a very compact starburst or an AGN torus (similar to those detected at high redshift by JWST, see \citealt{2023arXiv230801230M}, as well as \citealt{2023ApJ...957L...3P,2024arXiv240104159P}, and also \citealt{2017MNRAS.464.3431H}).



\subsubsection{Solar System origin}
\label{sec:solarsystem}

A bright mid-infrared object such as {\it Cerberus} matches the expected properties of Solar System bodies, e.g., asteroids. Given the long observational MIRI/F1000W campaign, we would expect to see proper motion, which is not the case, unless the asteroid is distant and/or near aphelion. However, the fact that the NIRCam data detect {\it Cerberus} (in stacked images) and the observations from MIDIS and JADES were taken more that one year apart indicates that {\it Cerberus} cannot be a Solar System body (assuming that the NIRCam and MIRI sources are the same object and not a fortuitous superposition). In Section~\ref{sec:validation}, we also measure the centroid shift of the object between the MIRI and NIRCam data to 0\farcs016 which is consistent with a zero offset. This makes the source even more unlikely to be a Solar System body.

\subsubsection{Galactic origin}

As shown by \citet{Langeroodi2023c}, \citet{2024ApJ...962..177B}, and \citet{2023arXiv230903250H} for LRDs, red sources detected by JWST in cosmological fields such as GOODS-S or A2744 can be brown dwarfs. Following the method in \cite{Langeroodi2023c}, we fit the photometry of {\it Cerberus} with brown dwarf atmosphere templates. We include the Sonora cloud-free models \citep{2021ApJ...923..269K, 2021ApJ...920...85M} as well as the ATMO2020++ cloud-free adiabat-adjusted T and Y dwarf models \citep[][see also \citealt{2022MNRAS.513.5701S}]{2021ApJ...918...11L, 2023AJ....166...57M}. In particular the ATMO2020++ models were included to ensure coverage of extremely cool brown dwarfs with 250 K $ < T_{\rm eff} <$ 500K. Compared to warmer brown dwarfs, these models more closely resemble the SED shape of {\it Cerberus}, which is relatively faint at near-infrared wavelengths while getting significantly brighter at 10~$\mu$m. 

The best-fit brown dwarf template is shown in the top-left panel of Figure \ref{fig:sps}. We infer a best-fit effective temperature ($T_{\rm eff}$ [K]) of $300^{+100}_{-25}$, surface gravity ($\log g$ [cm s$^{-2}$]) of $3.0^{+1.0}_{-0.5}$, and distance of $332 \pm 67$ pc. Its effective temperature would most likely classify {\it Cerberus} as a Y-type dwarf \citep[see Figure 12 in][]{Langeroodi2023c}. Compared to the UltracoolSheet compilation of known brown dwarfs with parallax distances \citep{ultracoolsheet}, its inferred distance ($\sim 300$pc) would make {\it Cerberus} one of the farthest (top 20) brown dwarfs discovered to date \citep[see Figure 11 in][]{Langeroodi2023c}. This distance at the high Galactic latitude of {\it Cerberus} would place it at the edge of the thin disk \citep[see Figure 6 in][]{2023arXiv230903250H}. Recently, three T-type dwarfs were discovered and spectroscopically confirmed with NIRSpec prism \citep{Langeroodi2023c, burgasser} at estimated distances between 0.7 and 4.8 kpc, all most likely outside the thin disk; among them A2744-BD3 is a late T-type at 755 pc. 


Our best-fit to a brown dwarf spectral template, shown in the top-left panel of Figure~\ref{fig:sps}, nicely fit the F444W$-$F1000W color and is consistent with the non-detection in the deep MIDIS F560W data. However, that brown dwarf model is very red in both the NIRCam SW-to-LW and 3-to-4~$\mu$m colors, which is not favored by our measurements (in stacked data), pointing to a flatter SED. 

In addition, both the MIRI and stacked NIRCam images (see Section~\ref{sec:validation}) shows the source to be slightly extended and asymmetrical as compared to the PSF. The two instruments, NIRCam and MIRI, indicate a morphology for the source that makes it unlikely that it is a point source, thus disfavoring the \textit{single} brown dwarf scenario.

The slight extension of order $0\farcs1$ alone would however not argue against a brown dwarf \textit{binary}. Even at only 100~pc distance, a semi-major axis of $0\farcs1$ for a binary of two identical brown dwarfs/planets with masses of $(0.0003-0.07)$~M$_\odot$ each would result in orbit periods of order 100--1000~years, i.e., the elongated image would not vary between the NIRCam and the MIRI observations.


\section{Summary and conclusions}
\label{sec:conclusions}

We report the discovery of a NIRCam-dark source identified in the MIRI Deep Imaging Survey (MIDIS) of the Hubble Ultra Deep Field carried out with the F1000W filter. We call this source {\it Cerberus}. The source has a magnitude around 27~mag (6$\sigma$) in F1000W and is extremely faint in all the NIRCam data taken by the deepest survey on the sky, the JWST Advanced Deep Extragalactic Survey, JADES, implying a magnitude fainter than $\sim$30.5 at wavelengths $\lesssim$5~$\mu$m. {\it Cerberus} is also undetected in the MIDIS F560W ultra-deep observations of the field, i.e., F560W$\gtrsim$29~mag. Our analysis of MIRI F1000W mosaics produced with limited datasets as well as the detection in NIRCam stacks at $\sim5\sigma$ level at wavelengths longer than $\sim3$~$\mu$m, both confirm {\it Cerberus} is real and qualifies as an extremely red object, with a color around F444W-F1000W $\sim$ 3.5~mag and F560W-F1000W $\gtrsim$ 2~mag. The morphology of the source in the MIRI F1000W image matches that observed in the NIRCam stacked data (with 1 year passing between both observations), in both cases presenting a main point-like component and similar-sized (i.e., unresolved) haze, further supporting that {\it Cerberus} is a real source not linked to a fortuitous alignment. 

We discuss the possible nature of {\it Cerberus} and identify 2 possible Galactic source types and 3 different kinds of galaxy that could match its spectral energy distribution. 

Given that the MIRI and NIRCam observations were taken with a one year epoch difference, we discard a Solar System origin. The NIRCam-to-NIRCam and NIRCam-to-MIRI colors do not match any known sub-stellar object, even the coolest and reddest brown dwarfs present mid-infrared colors that are not compatible with the spectral energy distribution of {\it Cerberus}.

Concerning an extragalactic origin, we identify 3 possibilities:

\begin{itemize}
\item A low-redshift solution, $z\sim0.4$, would imply that {\it Cerberus} is a dusty galaxy with strong emission from polycyclic aromatic hydrocarbons or warm dust heated by an active galactic nucleus. The low mass obtained for this solution, $\mathrm{M}_\star\sim10^5$~M$_\odot$, would point to an unknown type of dwarf galaxy with large dust content and no PAH depletion. 
\item The second possibility is a dusty starburst or post-starburst galaxy, or a galaxy hosting a mid-infrared bright obscured active galactic nucleus, at $z\sim4$ with a stellar mass $\mathrm{M}\sim10^8$~M$_\odot$ and high extinction.
\item The third possibility is a $z\sim15$ galaxy, the red color emanating from a spectral energy distribution dominated by emission from an obscured active nucleus with a bright torus significantly contributing to the flux in the rest-frame optical range, with shorter wavelengths dominated by a $\mathrm{M}_\star\sim10^{7}$~M$_\odot$ galaxy.
\end{itemize}

Despite vetting all the imaging data meticulously, confirming beyond doubt that the source is real, we have to accept that the available data are insufficient to establish the true nature of {\it Cerberus} with certainty. Given that the source is located in the deepest HST and JWST field, deeper imaging will likely be available in the near future and help to obtain a
better understanding of this unique object. 
In any case, further investigation of {\it Cerberus} will challenge the capabilities of JWST, as it was the case in the past for objects discovered at the limit of photometric surveys (remarkably, with space-based missions such as HST or {\it Spitzer}), whose spectroscopic follow-up (typically with ground-based facilities) was tremendously exigent (but still attempted with varying success). The discovery and analysis presented in this letter demonstrates the feasibility of identifying NIRCam-dark sources when selected as extremely red objects with MIRI observations up to at least 10~$\mu$m, maybe revealing new types of Milky Way sub-stellar bodies or of galaxy populations more easily detectable through ultra-deep photometric surveys in (some of) the reddest wavelengths probed by JWST.

 
\begin{acknowledgments}
We thank the referee for their constructive comments to our original manuscript. PGP-G and LC acknowledge support from grant PID2022-139567NB-I00 funded by Spanish Ministerio de Ciencia e Innovaci\'on MCIN/AEI/10.13039/501100011033,
FEDER {\it Una manera de hacer
Europa}.  KIC and EI acknowledge funding from the Netherlands Research School for Astronomy (NOVA). KIC acknowledges funding from the Dutch Research Council (NWO) through the award of the Vici Grant VI.C.212.036. MA acknowledges financial support from Comunidad de Madrid under Atracci\'on de Talento grant 2020-T2/TIC-19971.
This work was supported by research grants (VIL16599, VIL54489) from VILLUM FONDEN.
SG acknowledges financial support from the Villum Young Investigator grant 37440 and 13160 and the Cosmic Dawn Center (DAWN), funded by the Danish National Research Foundation (DNRF) under grant No. 140. JPP and TT acknowledge financial support from the UK Science and Technology Facilities Council, and the UK Space Agency. TPR acknowledges support from ERC Advanced Grant 743029 EASY. AAH acknowledges support from grant PID2021-124665NB-I00 funded by the Spanish
Ministry of Science and Innovation and the State Agency of Research
 MCIN/AEI/10.13039/501100011033 and ERDF A way of making Europe. JA-M, AC-G, LC acknowledge support by grant PIB2021-127718NB-100 from the Spanish Ministry of Science and Innovation/State Agency of Research MCIN/AEI/10.13039/501100011033 and by “ERDF A way of making Europe”. SEIB is funded by the Deutsche Forschungsgemeinschaft (DFG) under Emmy Noether grant number BO 5771/1-1.
JM, G\"O and AB acknowledge support from the Swedish National Space Administration (SNSA). RAM acknowledges support from the Swiss National Science Foundation (SNSF) through project grant 200020\_207349.

 Some of the data presented in this paper were obtained from the Mikulski Archive for Space Telescopes (MAST) at the Space Telescope Science Institute. The specific observations analyzed can be accessed via \dataset[DOI: 10.17909/je9x-d314]{https://doi.org/10.17909/wfg1-bm67}\dataset[DOI: 10.17909/je9x-d314]{https://doi.org/10.17909/je9x-d314}.
\end{acknowledgments}

\bibliography{f1000w_source_nircamdark}{}
\bibliographystyle{aasjournal}

\begin{deluxetable}{lllc}
\caption{Photometric measurements for the {\it Cerberus} source.}
\tablehead{\colhead{Band name} & \colhead{CWL} & \colhead{width} & \colhead{magnitude}
}
\startdata
F090W & 0.903 & 0.194 & $>30.0$ \\
F115W & 1.151 & 0.225 & $>30.4$ \\
F150W & 1.502 & 0.318 & $>30.9$ \\
F182M & 1.847 & 0.238 & $>29.8$ \\
F200W & 1.991 & 0.461 & $>30.0$ \\
F210M & 2.097 & 0.205 & $>30.1$ \\
F277W & 2.786 & 0.672 & $31.16\pm0.28$ \\
F335M & 3.364 & 0.347 & $30.10\pm0.35$ \\
F356W & 3.559 & 0.787 & $31.22\pm0.36$ \\
F410M & 4.084 & 0.436 & $30.39\pm0.40$ \\
F430M & 4.282 & 0.229 & $>30.2$ \\
F444W & 4.446 & 1.024 & $30.91\pm0.34$ \\
F460M & 4.630 & 0.228 & $29.69\pm0.36$ \\
F480M & 4.818 & 0.304 & $29.59\pm0.34$ \\
red\_stack\_01b & 4.497 & 0.755 & $29.89\pm0.20$ \\
red\_stack\_02b & 4.465 & 0.771 & $30.44\pm0.34$ \\
red\_stack\_03b & 4.386 & 0.665 & $30.43\pm0.29$ \\
red\_stack\_04b & 4.150 & 0.937 & $30.58\pm0.24$ \\
red\_stack\_05b & 4.068 & 1.047 & $30.51\pm0.17$ \\
red\_stack\_06b & 3.883 & 1.231 & $30.65\pm0.23$ \\
red\_stack\_07b & 3.791 & 1.296 & $30.84\pm0.19$ \\
red\_stack\_08b & 3.602 & 1.446 & $30.84\pm0.18$ \\
red\_stack\_09b & 3.520 & 1.524 & $30.95\pm0.24$ \\
blue\_stack\_01 & 1.049 & 0.392 & $>30.5$ \\
blue\_stack\_02 & 1.279 & 0.628 & $>30.7$ \\
blue\_stack\_03 & 1.438 & 0.846 & $>31.2$ \\
blue\_stack\_04 & 1.640 & 0.658 & $>30.7$ \\
blue\_stack\_05 & 1.703 & 0.742 & $>31.4$ \\
f2\_stack\_01   & 4.497 & 0.755 & $29.89\pm0.20$ \\
f2\_stack\_02   & 4.155 & 0.331 & $>30.1$ \\
f2\_stack\_03   & 3.514 & 0.602 & $30.58\pm0.19$ \\
f2\_stack\_04   & 2.596 & 0.750 & $>30.4$ \\
f2\_stack\_05   &  1.943 & 0.355 & $>30.5$ \\
f2\_stack\_06   &  1.380 & 0.496 & $>30.9$ \\
f3\_stack\_01   & 4.465 & 0.771 & $30.40\pm0.36$ \\
f3\_stack\_02   & 3.667 & 0.822 & $30.46\pm0.18$ \\
f3\_stack\_03   &  2.367 & 0.604 & $>31.0$ \\
f3\_stack\_04   &  1.536 & 0.711 & $>31.2$ \\ 
f4\_stack\_01   &  4.386 & 0.665 & $30.42\pm0.33$ \\
f4\_stack\_02   &  3.154 & 0.989 & $31.20\pm0.37$ \\
f4\_stack\_03   &  1.719 & 0.594 & $>31.1$ \\
f5\_stack\_01   &  4.150 & 0.937 & $30.50\pm0.32$ \\
f5\_stack\_02   &  2.467 & 0.892 & $>31.5$\\
F560W   &  5.645 & 1.000 &  $>29.7$ \\
F1000W  &  9.968 & 1.795 & $27.13\pm0.17$ \\
\enddata
\tablecomments{\label{tab:photometry}The table provides photometric measurements for {\it Cerberus} in all NIRCam individual images from JADES, incremental (named red and blue stacks) and disjoint (adding 2, 3, 4, and 5 contiguous filters) stacks, and MIRI bands from MIDIS. The central wavelengths and widths are given in $\mu$m, and the photometry in AB magnitudes, including $5\sigma$ upper limits for data points with $S/N<3$.}
\end{deluxetable}

\appendix
\section{Fitting codes used to analyze {\it Cerberus} spectral energy distribution}
\label{appA}

In this Appendix, we give details of the codes used to analyze the SED of {\it Cerberus} and obtain its possible redshifts if it were an extragalactic object, as well as the implications of those redshifts in its physical properties.

We first used the {\sc eazy} code \citep{2008ApJ...686.1503B} to fit the data in 2 ways: one using actual (low $S/N$)  flux measurements for all filters, and another replacing all $S/N<3$ measurements by $5\sigma$ upper limits that templates were not allowed to surpass (achieved with a modified version of the code). We employed in the fits all v1.3 templates for stellar-dominated galaxies \footnote{As listed here: \url{https://github.com/gbrammer/eazy-photoz/tree/master/templates}.}, which include a dusty galaxy with a high-equivalent width emission-line spectrum. We also allowed combinations of stellar-only templates with the new models for little red dots and high-redshift AGN$+$torus recently added based on JWST data \citep{2023arXiv231203065K}. We did not impose any prior and worked with minimum $\chi^2$ photometric redshift estimates in the range $0<z<20$. 

We also ran {\sc BAGPIPES} \citep{carnall2018}. {\sc BAGPIPES} is a stellar population synthesis modeling package built on the updated \cite{2003MNRAS.344.1000B} spectral library with the 2016 version of the MILES library \citep{miles}. It uses a \citet{kroupa2001} IMF. We adopted a \citet{2000ApJ...533..682C} dust attenuation allowing $0<\mathrm{A}_V<8$~mag, and included nebular emission lines. The SFH is set to delayed-$\tau$ model and the code included an AGN component as in \citet{carnall2023}.

We fitted the extracted photometry of {\it Cerberus} with {\sc prospector} \citep{prospector} to constrain its photometric redshift and stellar population properties. We adopted the {\sc prospector} setup described in detail in \cite{Langeroodi2023a} and \cite{Langeroodi2023b}. In brief, the SFH is modelled non-parametrically in five temporal bins with a continuity prior \citep[see e.g.,][]{2019ApJ...876....3L}; nebular emission is added from the {\sc cloudy} \citep{2023RMxAA..59..327C} runs compiled in \cite{2017ApJ...840...44B}; we treat the nebular- and stellar- metallicity as independent free parameters; dust attenuation is modelled with a two-component model, one for the entire galaxy and one for the star-forming regions \citep{2013ApJ...775L..16K}. In a first {\sc prospector} run, redshift is fitted as a free parameter with a flat prior in $0 < z < 20$. We then re-ran the same {\sc prospector} setup, fixing the redshift to the best-fit values mutually agreed using results from all the SED-fitting codes used in this work (see below).  

Complementarily, we made use of \textsc{LePHARE++}, the C++ latest evolution of the SED-fitting code \textsc{LePHARE} \citep{Arnouts2002,Ilbert2006B} to fit the {\it Cerberus} photometry, following the approach of Moutard et al. (2024, in prep.). In brief, we considered stellar population synthesis models from \cite{2003MNRAS.344.1000B} with a  \citet{2003PASP..115..763C} IMF and two metallicities ($Z_\odot$, 0.5 $Z_\odot$), assuming exponential SFHs with $0.1 \leq \tau  \leq 30$~Gyr, and delayed SFHs peaking after 1 and 3 Gyrs. To take into account the strong contribution from the nebular emission lines that can occur at very young ages, they are added following the line ratios adopted by \citet{Saito2020} and their normalization is allowed to vary by a factor of four. Aiming to take into account extremely dusty galaxies, attenuation is considered through the laws of \citet{2000ApJ...533..682C} and \citet{Arnouts2013}, with $0 \leq E(B-V) \leq 1.5$~mag, and dust IR reemission is taken into account following \citet{Bethermin2012A}. We finally ran \textsc{LePHARE++} on a grid between redshift $z=0$ and 20.




Apart from the previous codes, which were used to both constrain a photometric redshift and determine the physical properties of {\it Cerberus}, we also performed SED modeling with {\sc synthesizer-AGN} and {\sc cigale}, first fixing the redshift (to several values, see below).

The \textsc{synthesizer-AGN} code assumes that the SED can be modeled with a composite stellar population \citep{2003MNRAS.338..508P,2008ApJ...675..234P} and AGN emission coming from the accretion disk and the dust torus \citep{2024arXiv240108782P}. The stellar emission includes a  young and a more evolved star formation event, each one described by a delayed exponential function with timescales between 1~Myr and 1~Gyr, and with ages from 1~Myr up to the age of the Universe at the redshift of the source. The attenuations of the emission from each stellar population are independent and described by a \citet{2000ApJ...533..682C} law, with A$_{\rm V}$ values ranging from 0 to 10~mag for each population. The stellar emission is described by the \citet{2003MNRAS.344.1000B} models, assuming a \citet{2003PASP..115..763C} IMF with stellar mass limits between 0.1 and 100~$\mathrm{M}_\odot$, and the nebular emission is also considered (\citealt{2003MNRAS.338..508P}). The AGN emission is modeled with a QSO average spectrum 
\citep{2001AJ....122..549V,2006ApJ...640..579G}. The dust emission from the AGN is modeled with the self-consistent templates of AGN tori presented in \citet{2015A&A...583A.120S}, and dust emission linked to star formation is added using the models from \citet{2007ApJ...657..810D}.

CIGALE \citep{Boquien2019} SED-fitting analysis assume an SFH modelled using a constant star-formation rate with ages ranging from 1 to 100\,Myr. We adopted the stellar populations models from \citet{2003MNRAS.344.1000B} with solar metallicity, and the \citet{2003PASP..115..763C} IMF. We included nebular continuum and emission lines using solar metallicity, electron density of 100\,cm$^{-3}$, and ionised parameter equal to $\log(U)$=$-$2. The dust attenuation and far-IR emission uses the Calzetti Law \citep{2000ApJ...533..682C} and the \citet{Draine2014} models, respectively. AGN emission is added using the \citet{2006MNRAS.366..767F} models following the initial parameters suggested by \citet{Ciesla2015}.

The obtained photometric redshifts solutions are: $z\sim0.4$ (provided by {\sc prospector} and {\sc LePHARE++}), $z\sim4$ (very prominent solution for by {\sc eazy}, also detected weakly by {\sc prospector} and {\sc LePHARE++}), $z=14-15$ (obtained by all codes),  $z\sim17$ (obtained by {\sc LePHARE++}), and $z\sim19$ (obtained by {\sc bagpipes} and {\sc LePHARE++}). We note that the $z\sim0.4$ solution corresponds to a PAH line entering the F1000W passband, the $z\sim4$ solution would imply Pa$\alpha$ covered by the filter, for $z\sim15$, H$\alpha$ would contribute to the MIRI-F1000W flux, and $z\sim19$ means that the F1000W emission could be enhanced by H$\beta$ and/or [\ion{O}{3}]$\lambda\lambda$4960,5008 emission.


We fitted the SED shown in Figure~\ref{fig:sps}, which includes all flux data points with $S/N>3$ and upper limits for the rest of bands, using the codes described above. We checked that our results concerning the photometric redshift and physical properties (for the two parameters mentioned in the main text, the stellar mass and the dust attenuation) do not change significantly if we only use $S/N>5$ fluxes and upper limits, or just $S/N>5$ fluxes. More specifically, the photometric redshift solutions are the same, but with different statistical weights (i.e., the peaks vary their relative strength). Concerning the physical properties, variations were  very similar to the systematic differences between the distinct estimations for the two properties that are consistent among all codes and techniques, given that they are constrained by the two robust observational properties of {\it Cerberus}, i.e., the low luminosity and red color. More specifically, for the stellar mass we obtained variations of 0.3~dex when using higher signal-to-noise data for $z\sim4$ and $z\sim15$, and 0.7~dex for $z\sim0.4$. For the dust content, we obtained differences of $<0.3$~mag in all cases, always remaining relatively high, above 3~mag, or finding a similarly obscured torus model.

\end{document}